\documentclass[a4paper,11pt]{article}
\usepackage[utf8]{inputenc}

\usepackage[table]{xcolor}
\usepackage{jcappub}
\usepackage{graphicx}
\usepackage{psfrag,fancyhdr,epsfig}
\usepackage{hyperref}
\hypersetup{colorlinks,bookmarksopen,bookmarksnumbered,citecolor=magenta,linkcolor=blue,pdfstartview=FitH,urlcolor=blue}

\usepackage{cancel}
\usepackage{soul}

\usepackage{amsmath}
\usepackage{array}
\usepackage{subfig}
\usepackage{graphicx}
\usepackage{color}
\usepackage{tensor}
\usepackage{xcolor}
\usepackage{orcidlink}
\usepackage{comment}
\usepackage{physics}
\usepackage[mathscr]{eucal}
\usepackage{multirow}
\usepackage{graphicx}
\usepackage{amssymb}
\usepackage{amsmath}
\usepackage{subfig}
\usepackage{xcolor}

\newcommand{\be}{\begin{equation}}
\newcommand{\ee}{\end{equation}}
\newcommand{\bear}{\begin{array}}
\newcommand{\eear}{\end{array}}
\newcommand{\ba}{\begin{eqnarray}}
\newcommand{\ea}{\end{eqnarray}}

\usepackage{array}
\newcolumntype{C}[1]{>{\centering\arraybackslash}p{#1}}

\def\a{\alpha}
\def\b{\beta}

\usepackage[most]{tcolorbox}

\tcbset{colback=yellow!10!white, colframe=red!50!black, 
        highlight math style= {enhanced, 
            colframe=red,colback=red!10!white,boxsep=0pt}
        }

        \usepackage{empheq}
\newtcbox{\mymath}[1][]{%
    nobeforeafter, math upper, tcbox raise base,
    enhanced, colframe=blue!30!black,
    colback=blue!30, boxrule=1pt,
    #1}

\title{\huge \centering
Quasi-pole quintessential inflation\\ in metric-affine gravity }

\author[a]{Konstantinos Dimopoulos,}
\author[b,c]{Christian Dioguardi,\orcidlink{0000-0002-6133-0383}}
\author[b]{Ioannis~D.~Gialamas,\orcidlink{0000-0002-2957-5276}}
 \author[b]{Antonio Racioppi\orcidlink{0000-0003-4825-0941}}

\emailAdd{k.dimopoulos1@lancaster.ac.uk}
\emailAdd{christian.dioguardi@kbfi.ee}
\emailAdd{ioannis.gialamas@kbfi.ee}
\emailAdd{antonio.racioppi@kbfi.ee} 

\affiliation[a]{Consortium for Fundamental Physics, School of Physics and Astronomy,
Lancaster University, Lancaster LA1 4YB, United Kingdom}

\affiliation[b]{Laboratory of High Energy and Computational Physics, 
National Institute of Chemical Physics and Biophysics, R{\"a}vala pst.~10, Tallinn, 10143, Estonia}

\affiliation[c]{Tallinn University of Technology, Akadeemia tee 23, 12618 Tallinn, Estonia}

\abstract{We study quintessential inflation in the framework of metric-affine gravity. It is well known that non-minimal couplings with the Holst invariant can generate a quasi-pole inflationary behaviour resulting in a Starobinsky-like phenomenology. The same quasi-pole behaviour can also be used in order to ``flatten'' the scalar potential in the Dark Energy era providing a successful framework for quintessential inflation. Agreement with the observational constraints coming from bounds on the overproduction of gravitational waves and consistency with Big Bang Nucleosynthesis, reduces the predicted scalar spectral index to a narrow window when the non-minimal coupling to the Holst invariant is ``sufficiently" large:
$0.966 \lesssim n_s \lesssim 0.967$, 
making the model highly testable and falsifiable. 
}

\begin{document}

\maketitle

\section{Introduction}
\label{introduction}
Understanding the accelerated expansion of the Universe at both early~\cite{Kazanas:1980tx,Sato:1980yn,Guth:1980zm,Linde:1981mu} and late times is one of the central challenges of modern cosmology. The inflationary paradigm provides a compelling explanation for the remarkable homogeneity, isotropy, and flatness of the observable Universe, as well as the nearly scale-invariant spectrum of primordial density perturbations~\cite{Starobinsky:1979ty,Mukhanov:1981xt,Hawking:1982cz,Starobinsky:1982ee,Guth:1982ec,Bardeen:1983qw}. At late times, cosmic acceleration is attributed to Dark Energy, whose simplest realization is a cosmological constant, but requires incredible fine-tuning. This is why
dynamical models such as quintessence 
\cite{Caldwell:1997ii} have been put forward (For a review see Ref.~\cite{Copeland:2006wr}). Such proposals 
provide a richer and more flexible framework. Despite their successes, these two periods of accelerated expansion are typically described independently, leaving open the question of whether a unified explanation is possible. Moreover, dynamical Dark Energy, such as quintessence, requires an extra tuning, that of initial conditions.

Quintessential inflation gives a natural approach to this problem by using a single scalar field to drive both early- and late-time acceleration
\cite{Peebles:1998qn}. In this framework, the scalar field first dominates the energy density of the Universe during inflation and later reappears as a slowly evolving component that can account for Dark Energy (for recent reviews see Refs.~\cite{Jaman:2022bho,deHaro:2021swo}). This unified scenario has the potential to address the coincidence problem and to reduce the number of free parameters compared to separate models for inflation and quintessence while treating both phases of accelerated expansion in a common theoretical framework. It also explains the initial conditions of quintessence, which are determined by the inflationary attractor. Constructing viable quintessential inflation models, however, requires careful control over the scalar potential, as it must support two flat regions: the inflationary plateau and the quintessential tail, while also, allowing for a transition to a kination phase linking the two \cite{Joyce:1997fc}. Successful quintessential inflation models must account for the observations of both inflation and Dark Energy, without introducing too many parameters and without significant fine tuning. 

The gravitational framework in which the scalar field is embedded plays a crucial role in shaping the dynamics. Metric-affine gravity, where the metric and connection are treated as independent variables, provides a natural setting for introducing non-minimal couplings that can modify the effective scalar-field kinetic term and potential. In particular, couplings to curvature invariants such as the Holst invariant (see~\cite{Langvik:2020nrs,Shaposhnikov:2020gts,Piani:2022gon,Pradisi:2022nmh,Salvio:2022suk,Gialamas:2022xtt,Piani:2023aof,Gialamas:2023emn,DiMarco:2023ncs,He:2024wqv,Racioppi:2024zva,Inagaki:2024ltt,Gialamas:2024iyu,Racioppi:2024pno,Gialamas:2024uar,Karananas:2025xcv,Katsoulas:2025mcu,Gialamas:2025ciw,Salvio:2025izr,Gialamas:2025thp,Iosifidis:2025mcb,Karananas:2025fas,Karananas:2025qsm,Katsoulas:2025srh,Racioppi:2025igu,Gialamas:2026pjo} for recent cosmological applications in this framework) can generate quasi-pole structures in the kinetic function, stretching the potential in field space and producing plateaus suitable for inflation as well as for late-time acceleration. These features allow the construction of models that share some of the attractive properties of $\alpha$-attractor scenarios, while remaining consistent with quintessential inflation requirements \cite{Hossain:2014xha,Geng:2015fla,Dimopoulos:2017zvq,Dimopoulos:2017tud,Akrami:2017cir,Dimopoulos:2020pas,AresteSalo:2021wgb,Dimopoulos:2022tvn,Dimopoulos:2022rdp,Inagaki:2023mxv,Alho:2023pkl,SanchezLopez:2023ixx,Giare:2024sdl,Dimopoulos:2025fuq}.

The theoretical appeal of our quintessential framework is further reinforced by recent cosmological observations, which have revitalized the debate surrounding the nature of Dark Energy. While $\Lambda$CDM has enjoyed long-standing success, recent BAO measurements from the Dark Energy Spectroscopic Instrument (DESI) DR1~\cite{DESI:2024mwx} and DR2~\cite{DESI:2025zgx}, combined with CMB and SNe Ia data, exhibit a notable tension with the cosmological constant. Within the standard CPL parametrization~\cite{Chevallier:2000qy,Linder:2002et}, $w(a) = w_0 + w_a(1-a)$, these analyses yield a $3.1\sigma$ to $4.2\sigma$ preference for a dynamical Dark Energy (DDE) equation of state. Crucially, the best-fit parameters point toward a recent crossing of the ``phantom divide'' ($w=-1$), suggesting a transition from a phantom regime ($w < -1$) in the past to a quintessence regime ($w > -1$) today.

However, interpreting this phantom crossing as a physical reality warrants caution. From a theoretical standpoint, a true crossing in canonical minimally coupled single-scalar-field models is forbidden by the null energy condition, as it typically introduces ghost instabilities. From an observational perspective, the necessity of such a crossing is highly sensitive to the assumed priors. The rigid linear dependence of the CPL formulation can artificially force a crossing to accommodate low-redshift variations. As demonstrated in~\cite{Nesseris:2025lke}, extending the parameter space to higher-order CPL expansions or to quintessence like parametrizations~\cite{Gialamas:2024lyw,Gialamas:2025pwv} can significantly dilute the evidence for DDE, rendering the data consistent with $\Lambda$CDM. 

Consequently, single-field canonical quintessence remains an exceedingly viable and theoretically sound framework. The DESI data do not definitively mandate a violation of the null energy condition, but rather highlight the potential for a transient, dynamical equation of state with $w \ge -1$. This aligns perfectly with our model, which naturally yields a transient DDE phase as the field freezes on the potential plateau, avoiding the instabilities of phantom models. Thus, quintessence remains a robust alternative, justifying the detailed phenomenological exploration presented herein.

In this work, we present a model of quintessential inflation based on metric-affine gravity non-minimaly coupled to the Holst invariant. By appropriately choosing the coupling functions, the model naturally realizes a quasi-pole structure in the kinetic term, giving rise to plateau regions in the scalar potential that support both inflation and Dark Energy. The resulting inflationary predictions closely resemble those of Starobinsky-like models, while the late-time dynamics lead to a frozen or slowly rolling scalar field that can account for the observed Dark Energy density. The model therefore provides a minimal and predictive realization of quintessential inflation, bridging early- and late-time cosmic acceleration within a single coherent framework.

The paper is organized as follows. In section~\ref{the_model} we introduce the model, describing its main features and explaining how the two plateaus, one for inflation and one for Dark Energy, are generated through the Holst term. In section~\ref{sec:inflation} we introduce the well-known formalism for slow-roll inflation and compute the predictions for the CMB observables in our model. In section~\ref{sec:kination} we discuss kination, a period in which the scalar field kinetic term is the dominant source of the energy density of the Universe. This is a key feature of quintessential inflation models, that allows to put tighter bounds on the parameter space.
In section~\ref{sec:reheating} we briefly discuss reheating and explain how the reheating temperature is computed in our model.
In section~\ref{sec:quintessence} we describe the quintessential predictions of our model by studying the full evolution of the scalar field in presence of radiation and matter fluids.
In section~\ref{sec:parameters} we constraint the parameter space and highlight the predictions of our model.
Finally, in section~\ref{sec:conclusions} we discuss our results comparing it to the available results from Planck~\cite{Planck:2018jri,Planck:2018vyg}, Bicep/Keck~\cite{BICEP:2021xfz}, ACT~\cite{AtacamaCosmologyTelescope:2025nti} and DESI~\cite{DESI:2025zgx}.

\section{The model}
\label{the_model}

We start with the action
\be 
S = \int {\rm d}^4x\sqrt{-g}\left[\alpha(\phi){\cal R}+\beta(\phi)\tilde{\cal R}   - \frac12 (\partial_\mu \phi)^2 - V(\phi) \right], 
\label{eq:act1} 
\ee
where  $V(\phi)$ is the scalar potential, and $\alpha(\phi)$ and $\beta(\phi)$ are non-minimal coupling functions.  The quantities ${\cal R}$ and $\tilde{\cal R}$ represent, respectively, the Ricci scalar and the Holst invariant (i.e. the contraction of the Riemann curvature with the Levi-Civita antisymmetric tensor), both constructed from an independent connection $\Gamma$. Working in the metric-affine framework allows the connection to carry additional dynamics that can be integrated out at the classical level. Note also that the theory considered here, is dynamically equivalent to the Einstein-Cartan formulation, in which the connection is metric-compatible, $\nabla_\lambda g_{\mu\nu}=0$. This equivalence is a consequence of a projective symmetry of the action. It would be lost if one added terms constructed directly from the nonmetricity tensor  (see e.g.~\cite{Langvik:2020nrs}).

After integrating out the independent connection, the theory can be recast into the Einstein frame, where the gravitational sector is of the standard Einstein-Hilbert form. The resulting action reads (see e.g.~\cite{Langvik:2020nrs})
\be
\mathcal{S} = \int {\rm d}^4x \sqrt{-\bar{g}} \left[\frac{M_P^2}{2}R -\frac{k(\phi)}{2} (\partial_\mu \phi)^2 -U(\phi)  \right]\,,
\label{eq:act2}
\ee
where $M_P$ is the reduced Planck mass. All non trivial effects of the original non minimal couplings are now encoded in the non canonical kinetic function $k(\phi)$ and in the Einstein frame potential $U(\phi)$. These are given by
\be
k(\phi) =  \frac{M_P^2}{2} \left[\frac{1}{\a} + \frac{12(\b \, \partial_\phi\a -\a \, \partial_\phi\b)^2}{\a^2(\a^2+4\b^2)} \right]\,,  \qquad U(\phi)=\frac{V(\phi)}{F^2(\phi)} \, , \qquad
F(\phi) = \frac{2\a}{M_P^2}\,. \label{eq:k:U:F}
\ee
For the non-minimal couplings we adopt the choice\footnote{The careful reader might notice that a non-minimal coupling $\xi$ with the Ricci scalar could also be considered. In the following we assume that such a non-minimal coupling is small enough to be neglected in the numerical computations. More details are given in Section \ref{sec:parameters} in footnote \ref{footnote}.} 
\be 
\alpha(\phi) = \frac{M_P^2}{2}    , \qquad 
\beta(\phi) = \frac{M_P^2}{2} \left( \delta_{\beta}^2+ \tilde\xi \frac{\phi^2}{M_P^2} \right) \, ,
\label{eq:alphabeta}
\ee
which corresponds to working directly in the Einstein frame and implies $V(\phi)=U(\phi)$. Under this choice the kinetic function simplifies to
\be
k(\phi) = 1+ \frac{24\tilde{\xi}^2\phi^2 M_P^2}{M_P^4+4\left(\delta_{\beta}^2 M_P^2+\tilde{\xi}\phi^2\right)^2}\,.
\label{eq:kinetic}
\ee
 This function exhibits a non trivial structure controlled by $\delta_\beta$ and $\tilde\xi$. In particular, it possesses a local maximum located at
\be
 \phi_p^2 =  M_P^2 \frac{\sqrt{1+4 \delta _{\beta }^4}}{2 \left| \tilde{\xi }\right| } \, , \quad \text{with} \quad k(\phi_p)=1+6 |\tilde\xi|  \left(\sqrt{4 \delta _{\beta }^4+1}-\text{sign}(\tilde\xi) 2 \delta _{\beta }^2\right)\,.
 \label{eq:phi:max}
\ee
 which plays an important role in the dynamics. Although $k(\phi)$ does not develop true poles, for appropriate parameter choices the region around $\pm\phi_p$ can effectively mimic pole like behaviour~\cite{Racioppi:2024pno,Racioppi:2025igu}. From eq. \eqref{eq:phi:max}, we can see that the quasi-pole behaviour is favored when $\tilde\xi<0$, hence, from now on we will adopt such a choice.
Note that, after a field redefinition of the form $\sqrt{k(\phi)}{\rm d}\phi = {\rm d}\chi$, the kinetic term can be recast into canonical form. In terms of the canonical field, we denote by $\chi_p$ the position of the local maximum of the kinetic function $k(\phi)$.
We are interested in exponential potentials of the form
\begin{equation}
\label{eq:potin}
V(\phi) = V_0 e^{-\kappa\phi/M_P},
\end{equation}
where $\kappa$ is a dimensionless parameter controlling the slope and $V_0$ sets the overall energy scale and has dimensions of $\text{mass}^4$. Exponential potentials naturally arise in many high energy constructions\footnote{Ignoring the $D$-term contribution coming from eventual gauge fields, the scalar potential $V$ can be derived from the superpotential $W$ as $V = \sum_i \left| \frac{\partial W}{\partial \phi_i} \right|^2$ where $i$ runs over all the chiral superfields. Hence, if the superpotential is exponential, like in  \cite{Gorlich:2004qm,Haack:2006cy,Lalak:2005hr}, the scalar potential will be exponential as well. } \cite{Gorlich:2004qm,Haack:2006cy,Lalak:2005hr} and are particularly suitable for quintessential inflation scenarios~(e.g.~\cite{Jaman:2022bho,deHaro:2021swo}).

\begin{figure}[t]
\includegraphics[width=\textwidth]{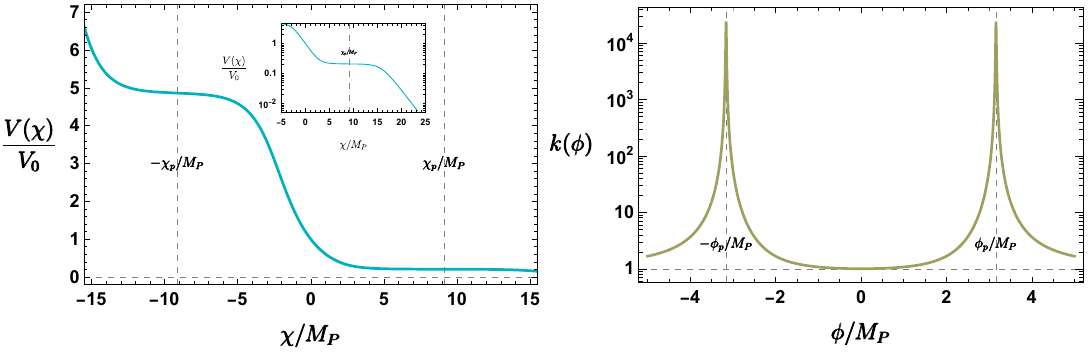}
\caption{\emph{Left}: The scalar field potential~\eqref{eq:potin} as a function of the canonical field $\chi$. The two plateau regions lie in the vicinity of the quasi-pole positions, indicated by the vertical dashed lines.
\emph{Right}: The kinetic function $k(\phi)$ given in eq.~\eqref{eq:kinetic} as a function of the scalar field.
In both panels, the parameters are $\delta_\beta = 10$, $\tilde{\xi} = -10$, and $\kappa = 1/2$.}
\label{fig:fig1}
\end{figure}

In figure~\ref{fig:fig1}, we show the potential as a function of the canonical field $\chi$ and the kinetic function as a function of $\phi$.  The effect of the quasi poles is clearly visible: near the positions of the quasi-poles the stretched field space flattens the potential, generating extended plateau regions. These plateaus can support slow-roll evolution even when the original potential in $\phi$ is steep, thereby opening the possibility for successful inflationary and Dark Energy dynamics within a unified setup.

\section{Inflation}
\label{sec:inflation}

The inflationary predictions of the model can be computed within the standard slow-roll approximation. After transforming to the canonical field $\chi$, the dynamics is entirely determined by the Einstein-frame potential $U(\chi)$, so we introduce the usual slow-roll parameters
\begin{equation}
    \epsilon(\chi) = \frac{M_P^2}{2}\left(\frac{U'(\chi)}{U(\chi)}\right)^2\,, \qquad
    \eta(\chi) = M_P^2 \frac{U''(\chi)}{U(\chi)}\,,
\end{equation}
 which quantify, respectively, the slope and curvature of the scalar potential.
 The duration of inflation is conveniently expressed through the number of $e$-folds,
\begin{equation}
    N_* = \frac{1}{M_P^2}\int^{\chi_*}_{\chi_{\rm end}}{\rm d}\chi \frac{U(\chi)}{U'(\chi)},
\end{equation}
where $\chi_{\rm end}$ denotes the field value at the end of inflation, determined by the condition $\epsilon(\chi_{\rm end})=1$. The value $\chi_*$ is instead computed by fixing the number of $e$-folds through~\cite{Liddle:2003as}:
\begin{align}  \label{eq:efolds}
N_* =& \ln \left(\frac{1}{\sqrt{3}}\left(\frac{\pi^2}{30}\right)^\frac{1}{4}\left(\frac{43}{11}\right)^\frac{1}{3}\frac{T_0}{H_0}\right) - \ln \left(\frac{k_*}{a_0 H_0}\right) - \frac{1}{12} \ln g_{\rm reh} \\ \nonumber
&+\frac{1}{4}\ln \left(\frac{U_*^2}{M^4_P \rho_{\rm end}}\right) + \frac{1-3w}{12(1+w)}\ln\left(\frac{\rho_{\rm reh}}{\rho_{\rm end}}\right),
\end{align}
and corresponds to the field value at the time at which the pivot scale leaves the horizon.
In \eqref{eq:efolds} $T_0$ is the temperature of the universe today, $H_0$ the Hubble constant, $k_*$ a pivot scale, $g_{\rm reh}$ the number of relativistic degrees of freedom at the time of reheating, $U_*$ the potential at the time when $k_*$ leaves the horizon, $\rho_{\rm end},\rho_{\rm reh}$ the energy density of the universe respectively at the end of inflation and at reheating.

 Within the slow-roll approximation, the inflationary observables are directly related to the slow-roll parameters. In particular, the tensor-to-scalar ratio $r$, the scalar spectral index $n_s$, and the amplitude of the scalar power spectrum $A_s$ are given by
\begin{equation}\label{eq:slow-roll}
r = 16\epsilon(\chi_*)\,,\quad   n_s= 1+2\eta(\chi_*) - 6\epsilon(\chi_*)\,, \quad A_s = \frac{1}{24\pi^2M_P^4}\frac{U(\chi_*)}{\epsilon(\chi_*)}\,.
\end{equation}
\begin{figure}[t]
 \centering
    {\includegraphics[width=1\textwidth]{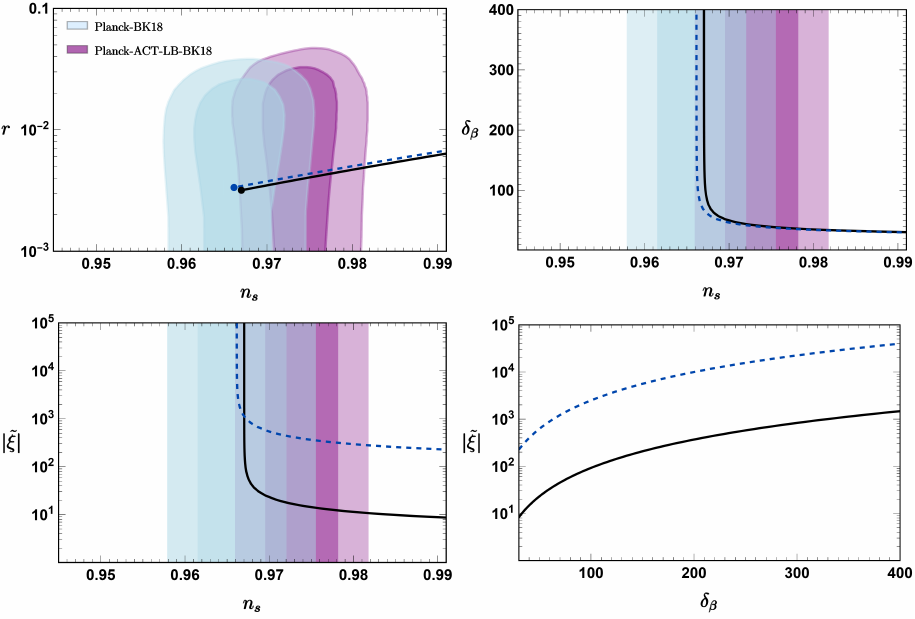}}%

\caption{$r$ vs.~$n_s$ (top-left), $\delta_\beta$ vs.~$n_s$ (top-right), $|\tilde\xi|$ vs.~$n_s$ (bottom-left), $|\tilde\xi|$ vs.~$\delta_\beta$ (bottom-right) for $\phi_p = 2$ at $N_* = 58.9$ (blue, dashed) and $\phi_p = 10.4$ at $N_* = 60.5$ (black, continuous). The plots are obtained numerically, using the slow-roll expressions \eqref{eq:slow-roll} and gradually increasing the parameter $\delta_\beta$ while keeping $\phi_p^2 \sim \delta_\beta^2/|\tilde{\xi}|$ fixed. The blue and black dots represent the predictions in the large $\gamma$ limit of eqs.~\eqref{eq:r:app}-\eqref{eq:As:app}, i.e.  the approximated Starobinsky attractor. Since these two models represent two extreme cases for our viable parameter space, the predictions for all other viable sets of $\phi_p$ and $N_*$ will fall between the two lines. See section \ref{sec:parameters} for the details on the parameter space, and section \ref{sec:conclusions} for the discussion of the results.}
    \label{fig:inflation}
\end{figure}
As shown in~\cite{Racioppi:2025igu}, models described by eq.~\eqref{eq:act1} exhibit an attractor behaviour. In particular, the inflationary predictions approach those of Starobinsky inflation independently of the specific form of $V(\phi)$, provided that the function $\beta(\phi)$ possesses a zero and is sufficiently steep in its vicinity.
Using eq.~\eqref{eq:alphabeta},  the attractor regime corresponds to $\delta_\beta \gg 1$, $\tilde\xi<0$ and $|\tilde\xi| \gg 1$. 
In this limit, the inflationary observables admit simple analytic approximations. Following~\cite{Racioppi:2025igu}, and working at large $N_*$, $\delta_\beta$, and $|\tilde\xi|$, we obtain
\begin{subequations}
\begin{eqnarray}
 r &\approx & \frac{12}{N_*^2}\left(1+ \frac{N_*^2}{27 {\gamma}^4} \right) \, , \label{eq:r:app} \\
 n_s &\approx & 1-\frac{2}{N_*}\left(1 - \frac{N_*^2}{27 {\gamma}^4} \right)  \, , \label{eq:ns:app}\\
 A_s &\approx & \frac{\lambda N_*^2 }{18 \pi ^2} \left(1-\frac{N_*^2}{27 \gamma ^4}\right) \label{eq:As:app}   \, ,
\end{eqnarray}
\end{subequations}
where we have defined
\begin{equation}
    \gamma \equiv -\frac{2 \delta_\beta \sqrt{|\tilde\xi|}    }{\kappa} \quad\text{and} \quad \lambda \equiv \frac{V_0}{M_P^4}  e^{\frac{\kappa \delta _{\beta }}{\sqrt{|\tilde\xi| }}}\,.
\end{equation}
At leading order in the large-$\gamma$ limit, these expressions reduce to the well-known predictions of Starobinsky inflation. The subleading corrections proportional to $\gamma^{-4}$ encode the residual dependence on the parameters of the model and quantify the departure from the exact attractor behaviour. As illustrated in Fig.~\ref{fig:inflation}, viable parameter choices cluster very close to the Starobinsky predictions, while still allowing for controlled deviations that can be probed observationally.

\section{Kination}
\label{sec:kination}

After inflation ends, the Universe naturally enters a period of kination, i.e. the kinetic energy of the scalar field becomes the dominant source of energy density of the universe. This period is model independent since the Klein-Gordon equation for the scalar field,
\be
\Ddot{\chi} + 3H\dot\chi + \partial_\chi U(\chi) = 0\,,
\ee
reduces to
\be
\Ddot{\chi} + 3H\dot \chi \simeq 0\,,
\ee
because the potential term becomes negligible compared to the Hubble friction and the acceleration terms \footnote{Notice that this trivially happens in the case of an exponential potential like \eqref{eq:potin} because $\partial_\chi U \propto U$ which becomes negligible right after the end of inflation.}.
The solution of this equation in terms the coordinate time $t$ is given by
\begin{equation}
    \dot\chi = \sqrt{\frac{2}{3}}\frac{1}{t}\,.
\end{equation}
As a consequence, the scalar energy density, $\rho_\chi$, evolves as
\be
\rho_\chi = \frac{1}{2}\dot\chi^2 + U(\chi)\sim \frac{1}{2}\dot\chi^2 \propto a^{-6},
\ee
where $a \propto t^{1/3}$ is the scale factor of the universe during the kination era. Comparing with the general scaling of the continuity equation, $\rho \propto a^{-3(1+w)}$, we deduce that the energy density during kination can be interpreted as the one of a fluid with barotropic parameter $w=1$. In our model we assume that all radiation is produced at the end of inflation through a mechanism that does not rely on inflaton decay, as we discuss in the next section.
Under this assumption, a period of kination generically interpolates between the end of inflation and the onset of radiation domination, which we identify as the reheating time. However, it is a well known fact that kination leads to an enhancement of primordial tensor modes that put a constraint on the duration of kination, $N_{\text{\rm kin}}$, itself. This constraint depends on the tensor-to-scalar ratio, and can therefore be relaxed for those models which provide lower values of $r$. Moreover, it was recently shown in~\cite{Eroncel:2025bcb} that kination also amplifies the well-measured amplitude of scalar modes leading to a stricter bound for $N_{\text{\rm kin}}$. Since our model predictions for inflation coincide with the Starobinsky one, we will use the upper bound for kination derived in~\cite{Eroncel:2025bcb} to constraint our parameter space (see section~\ref{sec:parameters} for the details).

\section{Reheating}
\label{sec:reheating}
In the quintessential inflation paradigm, the reheating process differs significantly from standard inflationary scenarios. Since the scalar field must survive to play the role of Dark Energy at late times, it cannot decay into Standard Model (SM) particles in the usual manner. Consequently, the radiation bath required for the Big Bang Nucleosynthesis (BBN) must be generated through alternative mechanisms, such as 
gravitational reheating~\cite{Ford:1986sy,Peebles:1998qn,Chun:2009yu,Hashiba:2018tbu,Haque:2022kez}, instant preheating~\cite{Felder:1998vq,Campos:2002yk,Dimopoulos:2017tud}, curvaton reheating \cite{Feng:2002nb,BuenoSanchez:2007jxm},
Ricci reheating~\cite{Dimopoulos:2018wfg,Opferkuch:2019zbd,Bettoni:2021zhq,Figueroa:2024asq,Bettoni:2024ixe}, reheating by primordial black hole evaporation \cite{Lennon:2017tqq,Martin:2019nuw,Dalianis:2021dbs,RiajulHaque:2023cqe}, warm quintessential inflation \cite{Dimopoulos:2019gpz,Rosa:2019jci}, to name but some. 

We define as ``reheating'' 
the moment when the energy density of radiation, $\rho_r$, becomes the dominating source of energy density in the Universe. For simplicity, we assume that the radiation density appears at the end of inflation. This is so in the case of gravitational reheating or warm quintessential inflation, when radiation is generated by particle production or by the diffusion of the inflaton kinetic energy density during inflation, but not afterward. Instant preheating also takes place at the end of inflation, because of the breakdown of adiabaticity when the inflaton field crosses an enhanced symmetry point, resulting into particle production at that moment. Finally, Ricci reheating operates also at the end of inflation.\footnote{
Ricci reheating considers the effect of a spectator field $\sigma$ non-minimally coupled to gravity as $\propto R\sigma^2$, whose effective mass squared flips sign at the end of inflation because \mbox{$R=3H^2(1-3w)$} with $w$ changing from $w=-1$ to $w=1$ from inflation to kination respectively. This displaces $\sigma$ from the origin and forces it to oscillate in a quartic potential, with energy density \mbox{$\rho_\sigma\propto a^{-4}$}, which decreases as radiation \cite{Opferkuch:2019zbd,Bettoni:2021zhq,Figueroa:2024asq,Bettoni:2024ixe}. 
The oscillating spectator field decays into relativistic particles at some point, but the density of the decay products continues to decrease as $\propto a^{-4}$. Therefore, for all intends and purposes, it does not really matter when the decay actually happens as long as it does before BBN. Thus, without loss of generality, we can assume that this decay occurs promptly at the end of inflation.}

The efficiency of the reheating process is characterized by the radiation energy density at the end of inflation, $\rho_r^{\rm end}$. In our framework, this is treated as a free parameter, effectively parameterized by the reheating temperature $T_{\rm reh}$. Assuming the produced radiation thermalizes sufficiently fast before the onset of the radiation-dominated era, the reheating temperature is given by
\cite{Dimopoulos:2017zvq}:
\begin{equation}
T_{\rm reh} = \left[\frac{30}{\pi^2 g_{\rm reh}} \left(\Omega_r^{\rm end}\right)^3 \rho_\chi^{\rm end} \right]^{1/4},
\label{eq:Treh}
\end{equation}
where $g_{\rm reh} = 106.75$ denotes the number of relativistic degrees of freedom in the SM at high temperatures, and $\Omega_r^{\rm end} = \rho_r^{\rm end}/\rho_{\rm tot}^{\rm end}$ represents the radiation density parameter at the end of inflation.

The duration of the kination phase, which occurs between the end of inflation and reheating, is crucial for the model's phenomenological viability. During kination, the scalar energy density redshifts as $\rho_\chi \propto a^{-6}$, while the radiation density redshifts more slowly as $\rho_r \propto a^{-4}$. As a result, lower values of $T_{\rm reh}$ (corresponding to lower $\Omega_r^{\rm end}$) imply a more protracted kination era. This period is subject to several observational constraints: (a) \textit{Scalar and tensor mode enhancement:} A prolonged kination phase can lead to an overproduction of primordial gravitational waves and an enhancement of the scalar power spectrum amplitude. We use the recent bounds derived in Ref.~\cite{Eroncel:2025bcb} for Starobinsky-like models to establish a lower bound on $T_{\rm reh}$. (b) \textit{BBN consistency:} To ensure the success of primordial nucleosynthesis, reheating must be completed well before the BBN epoch, imposing the absolute lower limit $T_{\rm reh} \gtrsim \mathcal{O}(1)$ MeV. (c)  \textit{Initial conditions:} We impose the consistency condition $\rho_\chi^{\rm end} > \rho_r^{\rm end}$ to ensure that a kination phase actually occurs, which provides a natural upper bound on the reheating temperature.

The specific viable window for $T_{\rm reh}$ in our model, obtained by numerically solving the evolution equations~\eqref{eq:friedmanN_evo}--\eqref{eq:matter_evo}, is discussed in detail in section~\ref{sec:parameters}.

\section{Quintessence}
\label{sec:quintessence}

To extract the quintessential predictions of our model, we numerically track the full background evolution of the scalar field in the presence of radiation and cold dark matter, using the number of $e$-folds $N$ as the time variable. The dynamics are governed by the first Friedmann equation alongside the continuity equations for the fluid components:
\begin{subequations}
\begin{align} 
\label{eq:friedmanN_evo}
  3  M_P^2 H^2 = & (H\chi')^2/2 + U(\chi) + \rho_r+\rho_m\,,
\\
  H^2 \chi''+ \left(3 H^2+H H'\right) \chi' = & -\partial_\chi U(\chi)\,,
\\
  \rho_r' + 4 \rho_r = & 0\,,
\\ 
  \rho_m' + 3 \rho_m = & 0\,,
\label{eq:matter_evo}
\end{align}
\end{subequations}
where $\rho_r, \rho_m$ are the energy densities of radiation and cold matter and the $'$ denotes the derivative with respect to $N$. The initial conditions are set under the assumption that the primordial radiation bath is generated at the end of inflation, while non-relativistic matter is produced during BBN. The initial abundance of radiation is left a priori unconstrained through the choice of $T_{\rm reh}$ or equivalently $\Omega_r^{\rm end}$. The upper bound on $\Omega_{r}^{\rm end}$ is set by considering the necessary condition $\rho_{\chi}^{\rm end}>\rho_{r}^{\rm end}$ at the end of inflation, while a lower bound can be obtained by considering the constraints coming from BBN observations (see section \ref{sec:parameters} for the details). The initial time (corresponding to $N=0$ in our simulation) is set at the time at which the pivot scale leaves the horizon during inflation.
As discussed in section~\ref{sec:reheating}, bounds on $T_{\rm reh}$ are self-consistently derived from the phenomenological limits on the duration of the kination epoch. 
Depending on the parameter space, the quintessential dynamics of our model split into two distinct phenomenological scenarios, illustrated in Figures~\ref{fig:benchmark_potential} and~\ref{fig:benchmark_2}.

In the \textbf{first scenario}, after inflation ends, the field rolls down the exponential potential and freezes during the radiation-dominated era \emph{before} reaching the late-time dark energy plateau. It remains frozen until its energy density becomes comparable to that of the dominant background fluid. Because the relation between $\phi$ and $\chi$ is approximately linear in this intermediate regime, the potential is effectively exponential. Consequently, the scalar field will follow the well-known scaling behavior of exponential quintessence. During this period the scalar field will mimic the barotropic parameter of the dominant fluid and its energy density will be a fixed fraction of the total, i.e.
\begin{align}
\label{eq:w_chi}
w_\chi &= w_B\,,
\\
\Omega_\chi &= \frac{3(1+w_B)}{\kappa^2}\,,
\label{eq:omega_chi}
\end{align}
where $w_\chi$ is the barotropic parameter of the scalar field, $w_B$ the barotropic parameter of the dominant background fluid, and $\Omega_\chi \equiv \rho_\chi/\rho_{\rm tot}$ is the fraction of the energy density of the scalar field to the total energy density of the universe.
After entering the scaling behavior the field will start evolving until it reaches the plateau. This will kick the scalar field out of the scaling behavior, allowing it to become the dominant source of energy density of the universe and behave as an effective cosmological constant. Reproducing the observed Dark Energy density requires an appropriate tuning of the potential parameters $V_0$ and $\kappa$. We provide a detailed benchmark for this scenario in section~\ref{sec:first_scenario}.

The \textbf{second scenario} exhibits qualitatively different late-time dynamics. Here, the field continues to roll down the exponential slope and does not freeze until it has already reached the plateau. Once there, it remains frozen with a nearly constant energy density until it becomes the dominant source of energy density in the Universe. Because the field is already situated on the flat plateau, it bypasses the scaling regime entirely, behaving as a cosmological constant, once it starts dominating. This case is further analyzed in section~\ref{sec:second_scenario}.

It is worth noting that in both scenarios, the predicted Dark Energy phase is transient. The finite length of the quintessential plateau in field space implies that, in the far future, the scalar field will eventually roll off the plateau. This will trigger a subsequent kination epoch ($\rho \propto a^{-6}$, $w=1$), which will ultimately give way to a second matter-dominated phase ($\rho \propto a^{-3}$, $w=0$).

\subsection{First scenario: field freezes before the Dark Energy plateau}\label{sec:first_scenario}

\begin{figure}[t]
    \centering
    {\includegraphics[width=1\textwidth]{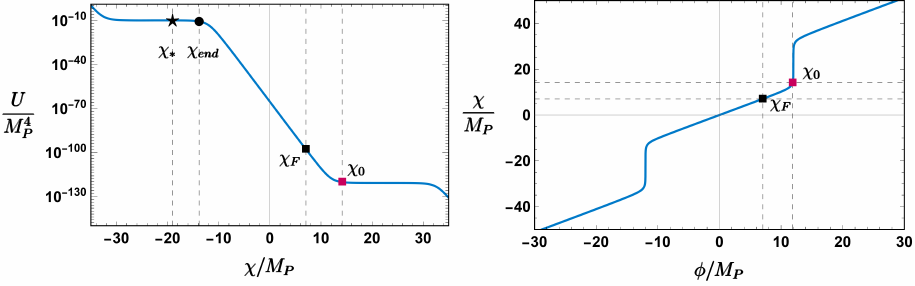}}%

   \caption{\emph{Left}: Benchmark potential for the first scenario with $|\phi_p| = 12$, $\delta_\beta=100$, $\tilde\xi =-69.5$, $\kappa=10.633$. The markers indicate critical field values: horizon exit of the pivot scale ($\chi_*$, black star), end of inflation ($\chi_{\rm end}$, black dot), freezing during radiation domination ($\chi_F$, black square), and present day ($\chi_0$, red square). \emph{Right}: The mapping between the canonical field $\chi$ and the original field $\phi$. Note that the field freezes while the $\chi$-$\phi$ relation is still linear.}
    \label{fig:benchmark_potential}
\end{figure}

\begin{figure}[t]
\includegraphics[width=1\textwidth]{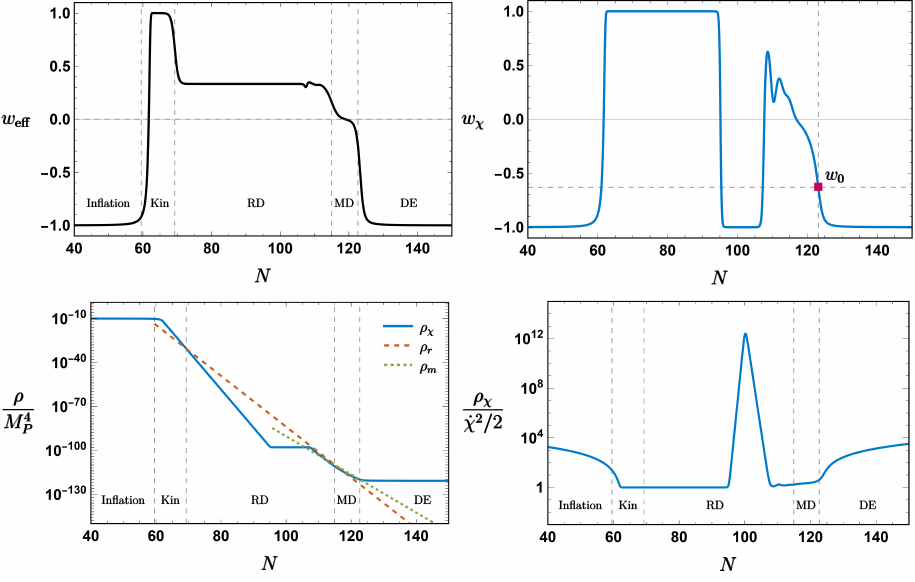}

\caption{\emph{Top left}: Evolution of the effective equation of state $w_{\text{eff}} \equiv P_{\rm tot}/\rho_{\rm tot}$ for the benchmark potential of Fig.~\ref{fig:benchmark_potential}. \emph{Bottom left}: Evolution of the energy density for the scalar field (blue), radiation (orange), and cold matter (green). \emph{Top right}: Barotropic parameter of the canonical scalar field. \emph{Bottom right}: Ratio of total to kinetic scalar energy density. The integration starts $N=40$ $e$-folds after horizon exit. Today corresponds to $N_0 \sim 123.1$.}
 \label{fig:scalar_evo}
 \end{figure}

We illustrate the first scenario using a benchmark point with $\delta_\beta = 100$, $\tilde\xi = -69.5$, and $\kappa =10.633$ (see Fig.~\ref{fig:benchmark_potential}, left). The pivot scale exits the horizon at $\chi_* \sim -19.04~M_P$. Inflation unfolds along the first plateau and ends at $\chi_{\rm end} \sim -13.84~M_P$, accumulating $N_* \sim 59.5$ $e$-folds. The corresponding CMB predictions are $r \sim 0.0033$ and $n_s \sim 0.9667$. Calibrating the amplitude of the scalar power spectrum to the observed value $A_s \sim 2.1\times 10^{-9}$ fixes the overall scale $V_0 \sim (108.9~\text{GeV})^4$, intriguingly close to the electroweak scale. As shown below, this value simultaneously resolves the Dark Energy coincidence problem.

Following inflation, a brief kination epoch take place, lasting $N_{\text{\rm kin}} \sim 9.8$ $e$-folds. This duration is perfectly compatible with the bounds derived in Ref.~\cite{Eroncel:2025bcb} for Starobinsky-like inflation. Assuming $g_{\rm reh}=106.75$, the associated reheating temperature is $T_{\rm reh} = 2.3 \times 10^{10}$ GeV corresponding to $\Omega^{\rm end}_r \sim 2 \times 10^{-7}$.
During radiation domination (RD), the scalar field depletes its kinetic energy and freezes at $\chi_F \sim 7.07~M_P$. As highlighted in Fig.~\ref{fig:benchmark_potential} (right), this freezing occurs while the relation between $\chi$ and $\phi$ is still strictly linear, meaning the Einstein-frame potential $U(\chi)$ remains purely exponential. Consequently, the field joins the subdominant scaling attractor, mimicking the background fluid without dominating the energy budget—a well-known hallmark of exponential quintessence. This scaling behavior is visible in Fig.~\ref{fig:scalar_evo} (bottom-left) near the end of the RD era. However, as the field slowly evolves and reaches the second plateau, the $\chi$-$\phi$ relation becomes non-linear. The presence of the plateau kicks the field out of the scaling regime, initiating the late-time accelerated expansion. At the present epoch ($N_0 \sim 123.1$), the scalar energy density is $\rho_\phi (N_0) \sim 7.15 \times 10^{-121}~M_P^4$, with barotropic parameter
$w_0 \sim -0.63$, consistent with Dark Energy constraints for dynamical models.

\subsubsection{A brief note on Early Dark Energy}

Equation~\eqref{eq:w_chi} dictates that the scaling attractor mimics the background fluid's barotropic parameter, while Eq.~\eqref{eq:omega_chi} shows that the field's fractional energy density remains constant during this regime. As the field enters the scaling solution, its energy density fluctuates. By appropriately tuning $\kappa$ (or $|\phi_p|$ via Eq.~\eqref{eq:parameter_kappa}) and $T_{\rm reh}$, it is possible to generate a localized peak in $\Omega_\chi$ near the epoch of matter-radiation equality. This phenomenological feature is characteristic of Early Dark Energy (EDE) models, which have been proposed to alleviate the $H_0$ tension when the peak amplitude reaches approximately $10\%$ of the total energy budget~\cite{Poulin:2023lkg}. 

\begin{figure}[t]
    \centering
    {\includegraphics[width=1\textwidth]{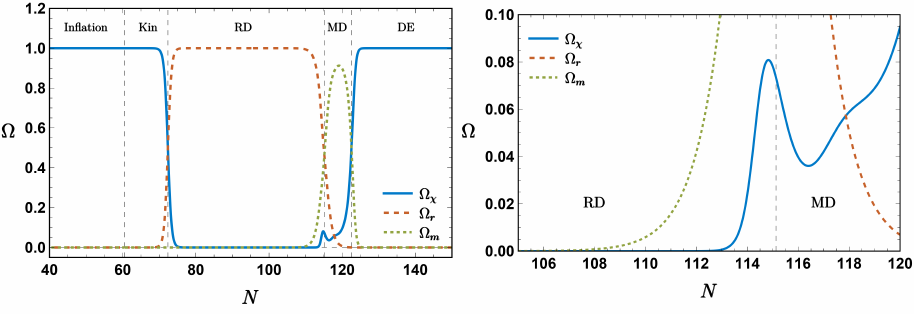}}%
 \caption{\emph{Left}: Evolution of the density parameters for the scalar field $\Omega_\chi$ (blue), radiation $\Omega_r$ (orange), and cold matter $\Omega_m$ (green). \emph{Right}: Magnified view around matter-radiation equality. The transient scaling behavior successfully generates an Early Dark Energy peak comprising $\sim 8\%$ of the total energy budget.}
    \label{fig:ede}
    
\end{figure}

Figure~\ref{fig:ede} demonstrates this mechanism for a parameter choice of $|\phi_p|=13.4$ and $T_{\rm reh} = 1.1 \times 10^9$ GeV. In this realization, the scalar field injects a transient energy density peak of $\sim 8\%$ near $N_{eq}$. While the peak amplitude is controlled primarily by $\kappa$, its temporal location depends sensitively on the initial conditions and $T_{\rm reh}$. A rigorous parameter estimation via MCMC to determine the viability of this specific EDE realization in resolving the Hubble tension is left for future work.

\subsection{Second scenario: field freezes on the Dark Energy plateau}\label{sec:second_scenario}

\begin{figure}[t]
    \centering
    {\includegraphics[width=1\textwidth]{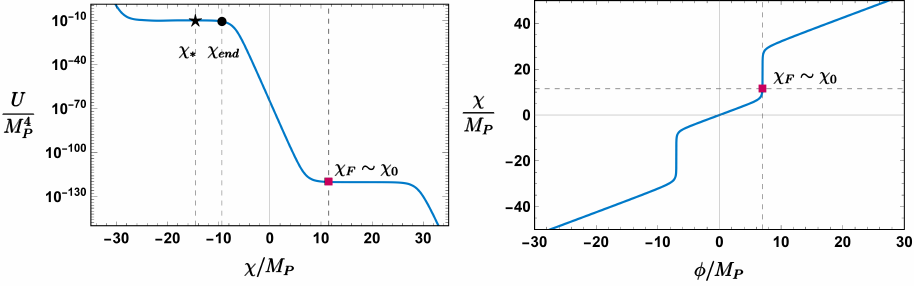}}%

 \caption{\emph{Left}: Benchmark potential for the second scenario with $|\phi_p|=7$, $\delta_\beta=100$, $\tilde\xi =-204$, and $\kappa=18.142$. Markers are the same with those in Fig.~\ref{fig:benchmark_potential}. \emph{Right}: The mapping between the canonical field $\chi$ and the original field $\phi$. Note that $\chi_F \sim \chi_0$.} 
    \label{fig:benchmark_2}
\end{figure}

\begin{figure}[t]
\includegraphics[width=1\textwidth]{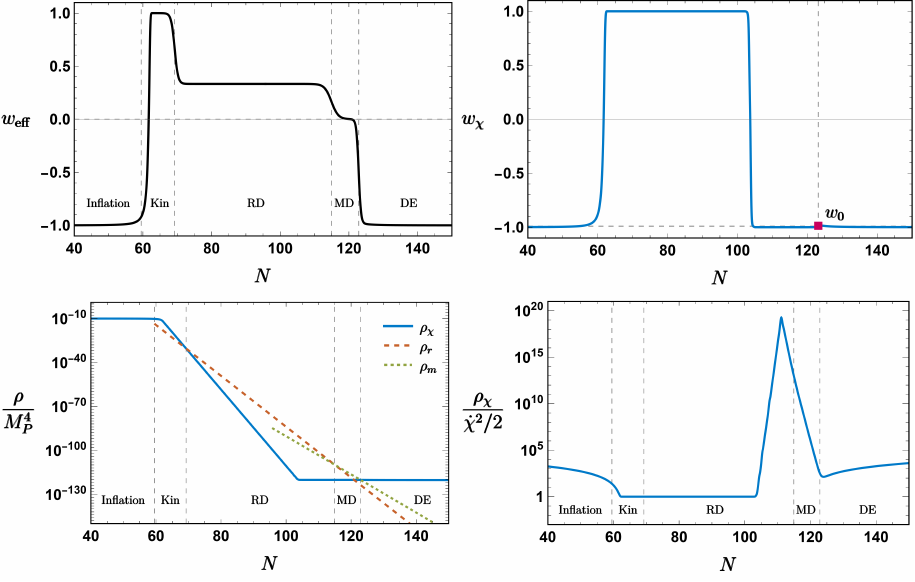}
\caption{\emph{Top left}: Evolution of $w_{\text{eff}}$ for the benchmark potential of Fig.~\ref{fig:benchmark_2}. \emph{Bottom left}: Energy densities. \emph{Top right}: Barotropic parameter of the canonical field. \emph{Bottom right}: Ratio of total to kinetic scalar energy density. The field skips the scaling regime entirely, remaining frozen on the plateau. The integration starts $N=40$ $e$-folds after horizon exit. Today corresponds to $N_0 \sim 123.1$.}
\label{fig:scalar_evo_2}
\end{figure}

In the second scenario, the field reaches the late-time plateau before freezing.  Figure~\ref{fig:benchmark_2} illustrates a benchmark point with $\delta_\beta = 100$, $\tilde{\xi} = -204.1$, $\kappa = 18.142$. Inflation occurs between $\chi^* \sim -14.56~M_P$ and $\chi_{\rm end} \sim -9.37~M_P$, yielding $N_* \sim 59.5$, $r \sim 0.0032$, and $n_s \sim 0.9667$.  $A_s$ is fixed to its observed value $A_s \sim 2.1\times 10^{-9}$ setting $V_0 \sim (127.1~\text{GeV})^4$, close to the EW scale. We will see that this value also manages to solve the coincidence problem for Dark Energy.

After the end of inflation the Universe enters a brief kination phase, lasting  $N_{\rm kin} \sim 9.7$ $e$-folds. Such a duration is fully compatible with the bounds on the kination epoch derived in Ref.~\cite{Eroncel:2025bcb} for Starobinsky-like inflation. Assuming $g_{\rm reh}=106.75$ relativistic degrees of freedom at reheating, the corresponding reheating temperature is
$T_{\rm reh} \sim 2.4\times 10^{10}\,{\rm GeV}$ corresponding to $\Omega^{\rm end}_r\sim 2 \times 10^{-7}$.
During the subsequent radiation-dominated era the scalar field gradually dissipates its kinetic energy and eventually freezes on the plateau of the potential at $\chi_F \sim 11.54\,M_P$. This represents a crucial difference with respect to the first scenario, where the field freezes before reaching the plateau. As shown in Fig.~\ref{fig:scalar_evo_2} (bottom-left), a direct consequence of this behavior is that the scalar field never enters a scaling regime and instead becomes the dominant contribution to the total energy density.
At late times the scalar field is therefore dominated by its potential energy on the plateau and effectively behaves as a cosmological constant. In particular, at $N_0 \sim 123.1$ we find $\rho_\phi(N_0) \sim 7.15\times 10^{-121}\,M_P^4$, consistent with the observed Dark Energy density, while the corresponding barotropic
parameter is $w_\chi(N_0) \equiv w_0 \sim -0.99$.

\section{Parameter space}\label{sec:parameters}

In this section we constraint the parameter space of our model, spanned by the ``Holst parameters" ($\delta_\beta, \tilde{\xi}$), the potential parameters ($V_0, \kappa$), the number of inflationary $e$-folds ($N_*$), and the reheating temperature ($T_{\rm reh}$). Using standard values for the pivot scale ($k_* = 0.05$ Mpc$^{-1}$), the Hubble parameter ($H_0 \sim 67.22$ km s$^{-1}$ Mpc$^{-1}$), the CMB temperature ($T_0 = 2.725$ K), and $g_{\rm reh} = 106.75$, Eq.~\eqref{eq:efolds} simplifies to

\begin{equation}
N_* \simeq 61.1 +\frac{1}{4}\ln \left(\frac{U_*^2}{M^4_P \rho_{\rm end}}\right) + \frac{1-3w}{12(1+w)}\ln\left(\frac{\rho_{\rm reh}}{\rho_{\rm end}}\right)\,.
\end{equation}
Assuming a purely kination-driven transition ($w=1$) between the end of inflation and reheating, the final term depends solely on the energy density ratios.

Figure~\ref{fig:inflation} displays the predictions for the CMB observables $r$ and $n_s$. In the strong-coupling limit ($|\tilde{\xi}| \gg 1$), the model accurately recovers the $\tilde{\xi}$-attractor behavior, continuously approaching the Starobinsky limit as $\delta_\beta \rightarrow \infty$. The black solid line corresponds to $\phi_p = 10.4$ at $N_* \sim 60.5$, yielding $T_{\rm reh} = 1.1\times 10^{9}$ GeV and $N_{\rm kin} \sim 11.8$. Conversely, the blue dashed line represents $\phi_p = 2$ at $N_* \sim 58.9$, with $T_{\rm reh} \sim 1.3\times 10^{11}$ GeV and $N_{\rm kin} \sim 10.1$. 

The phenomenological viability of these benchmarks is evaluated against recent constraints on the maximum duration of kination following Starobinsky-like inflation~\cite{Eroncel:2025bcb}, which restrict $N_{\rm kin} \lesssim 11.8$. Concurrently, a fundamental upper bound on $T_{\rm reh}$ arises from the necessary condition $\rho_\chi^{\rm end} > \rho_r^{\rm end}$. This implies $T_{\rm reh}^{\rm max} \sim 1.3\times 10^{11}$ GeV. 

All this induces the following constraints on the parameter space of our model (assuming all radiation is produced at the end of inflation):
\begin{equation}
N_{\rm kin} \lesssim 11.8\,,
\end{equation}
\begin{equation}
1.1\times 10^{9} \text{GeV} \lesssim T_{\rm reh} \lesssim 1.3 \times 10^{11} \text{GeV}\,,   
\end{equation}
\begin{equation}
3.4 \times 10^{-9} \lesssim \Omega^{\rm end}_r \lesssim 2 \times 10^{-6}\,,   
\end{equation}
way above the typical $\Omega^{\rm end}_{r,\rm grav}\sim 10^{-15}$ for gravitational reheating.
This correspondingly brackets the allowed number of inflationary $e$-folds to a narrow window (see eq.\eqref{eq:efolds}) given by
\be \label{eq:N_*}
58.9 \lesssim N_* \lesssim 60.5\,.
\ee
As depicted in Fig.~\ref{fig:inflation}, the lower bound for the parameter $\delta_\beta$ is independent of $\phi_p$ and hence independent of $\tilde{\xi}$ as well (this happens because all viable models are on the $\tilde\xi$-attractor, i.e. in the $|\tilde \xi| \gg 1$ limit). In particular compatibility with the $2\sigma$ region for $n_s$ requires $\delta_\beta \gtrsim 33$ (Planck-ACT-LB-BK18) or $\delta_\beta \gtrsim 38$ (Planck-BK18). The Starobinsky attractor is reached for $\delta_\beta \sim 70$. In this regime, given the narrow window~\eqref{eq:N_*} we obtain
\begin{equation}
r\sim 0.003\ \qquad \text{and} \qquad
    0.966 \lesssim n_s \lesssim 0.967\,.
\end{equation}
When $\delta_\beta, |\tilde{\xi}| \gg 1$, the relevant parameter for Dark Energy is given by $\phi_p^2 \sim \delta_\beta^2/|\tilde{\xi}|$. To each $|\phi_p|$ we have a fixed rather fine-tuned value of $\kappa$ necessary to match the observed value of the Dark Energy density today (smaller $\phi_p$ require larger $\kappa$). 
We can get an estimation for the potential parameters $\kappa, V_0$ by considering the following. The potential $U(\chi)$ in terms of the canonical scalar field has a plateau for $\chi \rightarrow \chi_p^-$, where inflation happens, and a plateau for $\chi \rightarrow \chi_p^+$, where we enter late time acceleration. In terms of $\phi$ this implies that:
\be
\begin{cases}
  U(\phi_p^-) = V_0 e^{-\kappa \phi_p^-} \sim 10^{-10} M_P^4\,,
\\ 
  U(\phi_p^+) = V_0 e^{-\kappa \phi_p^+} \sim 10^{-120} M_P^4\,,
\end{cases}
\ee
yielding
\begin{equation}
       V_0 \sim 10^{-65} M_P^4 \sim (100 \hspace{2pt}\text{GeV})^{4} \qquad \text{and} \qquad \kappa \sim 127/|\phi_p|,.
    \label{eq:parameter_kappa}
\end{equation}
While these provide excellent initial guesses, precision cosmology requires numerical integration of Eqs.~\eqref{eq:friedmanN_evo}--\eqref{eq:matter_evo}. Table~\ref{tab:parameters} summarizes the exact numerically obtained parameters in the Starobinsky limit.
\begin{table}\centering
\begin{tabular}{|C{1cm}||C{1cm}|C{1.2cm}|C{2cm}|C{1.2cm}|C{1.7cm}|C{1.2cm}|}
\hline
$\phi_p/M_P$ & $N_*$ &$N_{\rm kin}$& $\bar T_{\rm reh}\hspace{0.08cm}[\text{GeV}]$ & $\kappa$ & $V_0^{1/4}\hspace{0.08cm}[\text{GeV}]$ &$w_0$\\ 
\hline
$2.3$ & $58.9$ &$10.1$& $1.3\times 10^{11}$ & $55.305$ & $121.3$ &$-0.946$ \\ 
$5.1$ & $59.2$ &$9.8$& $5.5\times 10^{10}$ & $24.937$ & $121.3$ &$-0.951$\\ 
$8.8$ & $59.9$ &$10.5$ & $7.2\times 10^{9}$ & $14.453$ & $120.3$ &$-0.944$\\ 
$10.4$ & $60.5$ &$11.8$ & $1.1\times 10^{9}$ & $12.224$ & $121.4$ &$-0.958$\\ 
\hline
\end{tabular}
\caption{Model predictions and required potential parameters for selected values of $\phi_p$, computed in the Starobinsky limit ($\delta_\beta,|\tilde \xi| \gg 1$) enforcing $\rho_\chi(N_0)\sim 7.15\times 10^{-121} M_P^4$. Here, $\bar T_{\rm reh}$ denotes the maximum allowed reheating temperature compatible with $-1\lesssim w_0\lesssim -0.95$. Temperatures $T_{\rm reh} > \bar T_{\rm reh}$ push the dynamics into the scaling regime of the first scenario.}
\label{tab:parameters}
\end{table}

The parameter space fundamentally dictates the late-time phenomenology:\\

1) Models with $|\phi_p| \gtrsim 10.4~M_P$ configured for the second scenario require $N_{\text{\rm kin}} \gtrsim 11.8$, and are thus ruled out by the limits discussed in~\cite{Eroncel:2025bcb}. The limiting case $|\phi_p| \sim 10.4~M_P$ ($N_{\text{\rm kin}} \sim 11.8$, $T_{\rm reh} \sim 1.1\times 10^{9}$ GeV) allows exactly $N_* \sim 60.5$. Forcing more $e$-folds will imply a lower reheating temperature and larger $N_{\text{\rm kin}}$, which is excluded. Requiring a larger number of $e$-folds would lead to a lower reheating
temperature and a longer kination phase, thereby violating the above bound. Conversely, requiring fewer $e$-folds increases the reheating temperature and reduces $N_{\rm kin}$. In this case the scalar field $\chi$ freezes before reaching the plateau, which induces the scaling behaviour characteristic of the first scenario. Consequently, models with $|\phi_p|\gtrsim 10.4\,M_P$ predict
\begin{equation}
-0.95 \lesssim w_0\lesssim -0.63\,.   
\end{equation}
2) Models with $|\phi_p| \lesssim 2.3\,M_P$, on the other hand, naturally realize the second scenario (for an appropriate choice of $V_0$ and $\kappa$). Since the maximal reheating temperature is bounded by the inflationary plateau, $T_{\rm reh}$ can never become sufficiently large to trigger the first scenario. As a result, for these values of $\phi_p$ the present-day barotropic parameter lies in the range
 \begin{equation}
\label{eq:w0wa_pred}
-1\lesssim w_0\lesssim -0.95\,.
\end{equation}
3) Models with $2.3\,M_P \lesssim |\phi_p| \lesssim 10.4\,M_P$ admit a maximum reheating temperature $\bar T_{\rm reh}$ in the range 
$1.1\times 10^{9}\,{\rm GeV} \lesssim \bar T_{\rm reh} \lesssim 1.3\times 10^{11}\,{\rm GeV}$,
for which the scalar field freezes on the plateau, thereby realizing the second scenario. In these models the prediction 
$-1 < w_0 \lesssim -0.95$ is obtained only when
$1.1\times 10^{9}\,{\rm GeV} \lesssim T_{\rm reh} \leq \bar T_{\rm reh}$. 
Within this range the scalar field effectively behaves as a cosmological constant during the late-time accelerated expansion.
The precise value of $\bar T_{\rm reh}$ depends on $\phi_p$ and must be determined numerically by solving the full scalar-field evolution. In Table~\ref{tab:parameters} we present representative results for several values\footnote{\label{footnote} The value of $\phi_p$ allows us to determine how big the non-minimal coupling with the Ricci scalar $\xi$ can be, in order for it to be negligible. It is immediate to understand that such a bound is $\xi \ll M_P^2/\phi_p^2$. For instance, for the first row in table \ref{tab:parameters} we get $\xi \ll 0.19$, while for the last row we obtain $\xi \ll 0.009$. As mentioned before, an analysis for the case $\xi \gtrsim 1$ is postponed to a forthcoming work.} of $\phi_p$, together with the corresponding model parameters and predictions. All benchmark points are computed in the Starobinsky limit $\delta_\beta, |\tilde{\xi}| \gg 1$, imposing $\rho_\chi(N_0)\sim 7.15\times 10^{-121} M_P^4$.

\section{Conclusions}
\label{sec:conclusions}

In this paper, we considered a model of quintessential inflation embedded within the framework of Metric-Affine gravity. We demonstrated how non-minimal couplings to the Holst invariant naturally induce a quasi-pole structure in the effective kinetic term. Upon transforming to the Einstein frame, this geometric stretching generates two distinct, flat plateau regions on a single exponential potential, providing a minimal and theoretically well-motivated mechanism to drive both primordial inflation and late-time cosmic acceleration.

By imposing consistency conditions on the duration of the post-inflationary kination epoch and the onset of radiation domination, we strictly bounded the reheating temperature, restricting the allowed duration of inflation to $58.9 \lesssim N_* \lesssim 60.5$. In the strong-coupling Starobinsky limit ($\delta_\beta \gtrsim 70$), this tightly constrains the scalar spectral index to the narrow, highly predictive window $0.966 \lesssim n_s \lesssim 0.967$. This prediction is in excellent agreement with the Planck-BK18 data~\cite{BICEP:2021xfz}. While there is a mild $\sim 2\sigma$ tension with the joint Planck-ACT-LB-BK18 dataset~\cite{AtacamaCosmologyTelescope:2025nti}, we note that this discrepancy primarily stems from the inclusion of DESI BAO data~\cite{DESI:2025zgx}, which introduces a known BAO-CMB tension independent of inflationary physics~\cite{Ferreira:2025lrd}. Furthermore, alignment with the ACT data can still be achieved outside the strict Starobinsky limit by adopting intermediate coupling values ($35 \lesssim \delta_\beta \lesssim 50$) in which case the model provides a wider window: $0.966 \lesssim n_s  \lesssim 0.981$.

At late times, our numerical analysis revealed two viable quintessential regimes. Depending on the geometry of the plateau ($\phi_p$) and the thermal history ($T_{\rm reh}$), the scalar field can either transiently follow a scaling attractor before reaching the Dark Energy plateau (predicting an equation of state with barotropic parameter $-0.95 \lesssim w_0 \lesssim -0.63$, with potential implications for Early Dark Energy), or freeze directly on the plateau, effectively mimicking a pure cosmological constant with $-1 < w_0 \lesssim -0.95$. In both limits, the framework successfully addresses the coincidence problem using natural mass scales ($V_0 \sim \text{TeV}^4$) while unifying the early and late accelerating phases of the Universe.

\acknowledgments

KD was supported (in part) by the Consortium for Fundamental Physics under STFC consolidated grant: ST/X000621/1.
This work was supported by the Estonian Research Council grants PSG1132, PRG1677, TARISTU24-TK10, TARISTU24-TK3, and by the CoE program TK202 ``Foundations of the Universe''. This article is based upon work from COST Action CosmoVerse CA21136, supported by COST (European Cooperation in Science and Technology).

\bibliography{quint_bib}{}
\end{document}